\documentclass[12pt]{article}
\usepackage[english]{babel}
\usepackage{graphicx}
\usepackage{lineno}
\usepackage{hyperref}
\usepackage{titlesec}
\titlespacing*{\section}{0pt}{0.9\baselineskip}{0.5\baselineskip}

\newcommand\pubdate{\today}

\def\Title#1{\begin{center} {\Large #1 } \end{center}}
\def\Author#1{\begin{center}{ \sc #1} \end{center}}
\def\Address#1{\begin{center}{ \it #1} \end{center}}

\newcommand\pubblock{\rightline{\begin{tabular}{l}  \\ % Author's note number [if you need to add one] goes here
         \pubdate  \end{tabular}}}
\newenvironment{Abstract}{\begin{quotation}  }{\end{quotation}}
\newenvironment{Presented}{\begin{quotation} \begin{center}
             PRESENTED AT\end{center}\bigskip
      \begin{center}\begin{large}}{\end{large}\end{center} \end{quotation}}

\renewenvironment{thebibliography}[1]
  {\begin{list}{\arabic{enumi}.}
  {\usecounter{enumi}\setlength{\parsep}{0pt}
      \setlength{\itemsep}{0pt} 
      \settowidth
  {\labelwidth}{#1.}\sloppy}}{\end{list}}
\def\Journal#1#2#3#4{{#1} {\bf #2}, #3 (#4)}

\def\PRD{{\em Phys. Rev.} D}

\def\EPJC{{\em \it Eur. Phys. J.} C}

\begin{document}
  %\linenumbers
  \begin{titlepage}
    \pubblock
    \vfill
    \Title{Searches for supersymmetric particles with prompt decays with the ATLAS detector}
    \vfill
    \Author{Francesco Giuseppe Gravili, on behalf of the ATLAS Collaboration}
    \Address{University of Salento, Department of Mathematics and Physics ``Ennio De Giorgi'', \\ \& INFN Unit of Lecce, 73100 Lecce, Italy}
    \vfill
    \begin{Abstract}
      Supersymmetry (SUSY) provides elegant solutions to several problems in the Standard Model and searches for SUSY particles are an important component of the LHC
      physics program. The latest results from electroweak and strong SUSY searches are reported here, conducted by the ATLAS experiment at the CERN LHC. 
      The searches target multiple final states and
      different assumptions about the decay mode of the produced SUSY particles, including searches for both R-parity conserving models and R-parity violating
      models, and their possible connections with the recent observation of the flavour and muon \textit{g-2} anomalies. The talk will also highlight the employment of
      novel analysis techniques, including advanced machine learning techniques and special object reconstruction, that are necessary for many of these analyses to
      extend the sensitivity reach to challenging regions of the phase space.
    \end{Abstract}
    \vfill
    \begin{Presented}
      DIS2023: XXX International Workshop on Deep-Inelastic Scattering and Related Subjects, \\
      Michigan State University, USA, 27-31 March 2023\\
      \includegraphics[width=9cm]{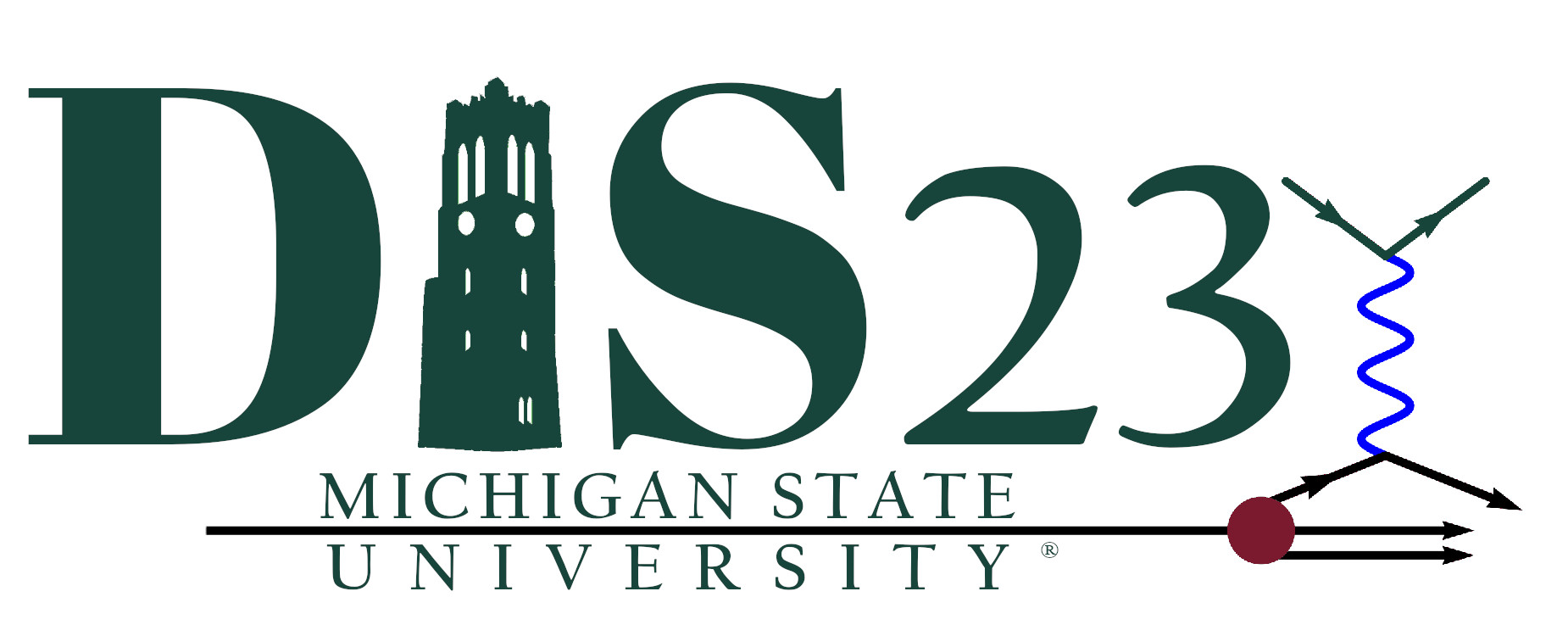}
    \end{Presented}
    \vfill
    Copyright 2023 CERN for the benefit of the ATLAS Collaboration. CC-BY-4.0 license
  \end{titlepage}

  \section{Introduction}
  The Standard Model (SM) of fundamental interactions is the underlying theory of elementary particles and their interactions. 
  Despite the many experimental results precisely confirming the predictions of this theory, there are still some open questions within the model.
  SUperSYmmetry (SUSY) is a Beyond Standard Model (BSM) physics theory, predicting the existence of fermionic (bosonic) supersymmetric partners 
  for the bosons (fermions) in the SM, differing by $1/2$ unit in spin. 
  A new quantum number is introduced, \textit{R}-parity, $R = (-1)^{3(B-L)+2s}$ with \textit{B}, \textit{L} being the
  baryon and lepton numbers, while \textit{s} is the spin of the particle. According to the violation or not of this quantum number, two scenarios are then introduced: \textit{R}-parity-violating (RPV) and \textit{R}-parity-conserving (RPC) models.
  The superpartners of the SM Higgs (\textit{higgsinos}) and electroweak (EWK) gauge bosons (collectively referred to as \textit{electroweakinos}) 
  mix to form \textit{chargino} ($\tilde{\chi}_i^{\pm}$, $i=1,2$) and \textit{neutralino} ($\tilde{\chi}_j^0$, $j=1,\dots,4$) mass eigenstates. 
  The $\tilde{\chi}_1^0$ is usually assumed to be the lightest supersymmetric particle (LSP) in the SUSY decay chains.
  \newline
  The latest results from ATLAS SUSY searches are highlighted, both for \textit{R}-parity conserving and \textit{R}-parity violating models.
  Several final states are targeted, in the context of SUSY prompt decays, i.e. $c \tau < \mathcal{O}(1)$ mm. Searches exploit the full LHC Run 2 proton-proton collision 
  dataset collected with the ATLAS~\cite{atlas} detector at the CERN LHC, at $\sqrt{s}=13$ TeV and corresponding to an integrated luminosity of $139$ fb$^{-1}$.

  \section{\textit{R}-parity-violating searches}
  A BSM search for pair production of supersymmetric particles with RPV decays into final states with high jet multiplicity, 
  at least one isolated light lepton and either zero or at least three \textit{b}-tagged jets is presented~\cite{RPVmulti-j}. The RPV analysis models involve either 
  baryon-number-violanting $\lambda^{'''}_{323}$ coupling and lepton-number-violating $\lambda^{'}$ coupling. 
  $\lambda^{'''}_{323}$ is assumed to be dominant and similar final states apply to $\lambda^{''}_{313}$ as well.
  With this choice of model parameters, final state signatures includes $\tilde{\chi}^0_{1/2} \rightarrow tbs$ and $\tilde{\chi}^{\pm}_1 \rightarrow bbs$ 
  with a Branching Ratio (BR) of 100\% and $\tilde{\chi}_1^0 \rightarrow q \bar{q} \ell / \nu$, with an equal probability to produce any of the four first- and 
  second-generation leptons.
  Events are splitted into two categories, according to the lepton content: two same electric charge leptons (SS) and single lepton. 
  Due to the presence of many jets in the final states, a multi-bin fit is performed in each lepton category and a key variable, based on a machine-learning
  discriminant, is introduced to improve the sensitivity. No significant excess is observed over SM expectations, and results are interpreted in the framework 
  of simplified models. In the model with $\tilde{g} \rightarrow t\bar{t} \tilde{\chi}_1^0 \rightarrow t \bar{t} tbs$, gluino masses up to 2.38 TeV 
  are excluded at 95\% confidence level (CL). Top squarks masses up to 1.36 TeV are excluded in a model with direct stop production and RPV decays of the LSP. 
  Concerning the direct production of electroweakinos, higgsino (wino) masses between 200 (197) GeV and 320 (365) GeV are excluded.
  \medskip
  \newline
  The introduction of a bilinear lepton-number-violating term allows to test sensitivity to the direct pair production of light, nearly mass degenerate higgsinos,
  $\tilde{\chi}_2^0$, $\tilde{\chi}^{\pm}_1$ and $\tilde{\chi}_1^0$. Mass splittings are below 2 GeV.
  The dominant production processes feature same electric charge sign dilepton and trilepton final states. The inclusive production and all possible 
  allowed higgsino decays are considered in this analysis~\cite{bRPV}.
  Observed data are compatible with SM predictions, and limits are set on the parameters for this scenario: after a statistical combination of the two orthogonal 
  signal regions (SRs), one for each final state, mass degenerate higgsinos are excluded up to masses of 440 GeV. 
  These are the first experimental constraints on bilinear RPV (bRPV) models with degenerate higgsino masses.

  \section{\textit{R}-parity-conserving searches: \textit{Strong} sector}
  A search for new phenomena in final states with one or more hadronically decaying $\tau$-leptons, \textit{b}-jets and missing transverse momentum is 
  presented~\cite{LQ}. This signature is sensitive to models in which the new particles have preferential decay modes into third-generation SM particles
  and leptoquarks (LQ). In the latter case, the analysis is optimized for scalar LQ, while an additional 
  interpretation is provided for vector LQ. The analysis covers the single-$\tau$ and di-$\tau$ channels separately. Hadronically decaying $\tau$-leptons 
  are distinguished from quark- and gluon- initiated jets thanks to a recurrent neural network. A combination of high-level discriminating variables, 
  as well as tracking and calorimeter measurements, is given as input. In the case of the supersymmetric model, masses up to 1.4 TeV are excluded for top squarks 
  decaying via $\tilde{\tau}$ sleptons into nearly massless gravitinos $\tilde{G}$,
  across a wide range of $\tilde{\tau}$ masses. In the case of \textit{up}-type and \textit{down}-type scalar LQ, masses up to about 1.25 TeV are excluded. 
  On the other hand, for vector LQ, masses up to about 1.8 TeV are excluded.
  \medskip
  \newline
  Another key search involves the direct pair production of gluinos decaying via off-shell third-generation squarks into the lightest neutralino~\cite{multi-b}.
  The final state signature is multiple \textit{b}-jets and high missing transverse momentum; potentially, additional jets and/or an isolated electron or muon. 
  Three different benchmark simplified models scenarios are introduced: the first two, referred to as \textit{Gtt} and \textit{Gbb}, 
  feature exclusively gluino decays to the LSP via off-shell top or bottom quarks. The last one, referred to as \textit{Gtb}, takes into account
  different branching ratios for the $t \bar{t}$, $b \bar{b}$, $t \bar{b}$ and $b \bar{t}$ decays. Two alternative methodologies are used to define SRs:
  for \textit{Gtb} models, a standard cut-and-count approach is used, well suited to subsequent reinterpretation of the results. 
  For \textit{Gtt} and \textit{Gbb} models, a neural network methodology classifies events into 4 different output scores (\textit{Gtt} or 
  \textit{Gbb} signal event, $t \bar{t}$ or \textit{Z}+jets background event), 
  exploiting correlations between the input discriminating variables to maximise the exclusion power.
  No significant excess over the expected SM predictions is observed, and exclusion limits are set on the SUSY particles involved in the models: 
  for the exclusively \textit{Gtt} and \textit{Gbb} gluino decays, masses below 2.44 and 2.35 TeV are excluded at 95\% CL for massless neutralinos, respectively.
  \medskip
  \newline
  Other exclusion limits on gluino masses are coming from a general gauge mediated (GGM) analysis, featuring many jets in the final states, 
  in combination with a highly energetic photon and a gravitino. It is considered as the LSP, coming from the decay of a next-to-lightest 
  SUSY particle (NLSP), typically the lightest neutralino~\cite{photon}. The decay topologies $\gamma/Z$ and $\gamma/h$ are targeted in the analysis, 
  with the first ATLAS Run 2 results for the latter one. Good agreement is found between observed data and expected SM backgrounds in all the SRs, and pair-produced 
  gluinos with masses up to 2.2 TeV are excluded for most of the NLSP masses investigated.

  \section{\textit{R}-parity-conserving searches: \textit{EWK} sector}
  In the context of gauge mediated supersymmetry models (GMSM), one of the latest ATLAS result is about the direct pair production of higgsinos, 
  decaying into a light gravitino either via a Higgs or \textit{Z} boson~\cite{HPhot}. The final state features a photon pair coming from the Higgs boson, 
  a $b \bar{b}$ pair coming from the other Higgs or \textit{Z} boson and missing transverse momentum associated with the two gravitinos. 
  Events are required to pass diphoton triggers and three different SRs are defined. This choice allows to gain sensitivity
  to different mass hypoteses and decay modes, differing in the requirements on the invariant mass of the $b \bar{b}$ pair and on the missing transverse momentum. The
  first two SRs target small higgsino mass ranges, with the invariant mass requirement being consistent to the Higgs/\textit{Z} boson mass.
  The last SR is designed for higher mass higgsino decays, consequently demanding high missing transverse momentum. 
  SR observed yields are found to be consistent with SM expectations, and exclusion limits are set on pure-higgsino branching ratio
  $BR(\tilde{\chi} \rightarrow h\tilde{G})$ against the higgsino mass, assuming the two aforementioned decay topologies. This analysis fills 
  the gap left by previous analysis signatures in that phase space. A statistical combination of the SRs is performed to derive limits on cross section for 
  higgsino pair production: cross sections above 1 pb are excluded at 95\% CL for masses higher than 150 GeV, and the theoretical prediction for the pure higgsino
  cross section is excluded at 95\% CL for neutralino masses below 320 GeV.
  \medskip
  \newline
  In terms of complementary analyses, the search for electroweak production of chargino pairs with decays into a \textit{W} boson 
  (both leptonic and hadronic consequent decays are allowed) and the lightest neutralino is presented~\cite{EWK1L}. 
  Single lepton triggers are used to identify candidate events. The requirement of at least one \textit{large-Radius} jet allows to probe 
  boosted \textit{W} boson hadronic decays.
  Three SRs are defined, using the transverse mass to target regions sensitive to the increasing mass difference between the lightest chargino and neutralino,
  making them mutually exclusive. No significant deviation from the SM expectations are observed in any of the SRs, and chargino masses between 260 and 520 GeV
  are excluded for massless neutralino. Previous ATLAS searches targeted the low- and high-mass areas, while the current one covers the intermediate region.
  \medskip
  \newline
  Concerning sleptons, the direct pair production of electroweakinos decaying via intermediate $\tilde{\tau}$ sleptons into final states with 
  hadronically decaying $\tau$-leptons is presented~\cite{staus}. Following a standard cut-and-count based approach, using kinematic variables with good
  signal-to-background separation, exclusion limits are extended for high $\tilde{\chi}^{\pm}_1/\tilde{\chi}^0_2$ degenerate masses up to 1160 GeV for massless 
  lightest neutralino. Previous results are improved by 340-400 GeV.
  Finally, a general overview of SUSY results for the direct $\tilde{\mu}$ pair production is presented. Interesting regions of phase space are highlighted, 
  consistent with \textit{g-2} anomaly under different assumptions of SUSY parameters, 
  as well as possible future strategies in order to tackle the $\tilde{\mu}$-corridor~\cite{summary}.

  \section{Summary}
  The latest ATLAS searches for supersymmetric particles with prompt decays are described, both for RPV and RPC scenarios. 
  The searches used proton-proton collision data at $\sqrt{s}=13$ TeV, corresponding to an integrated luminosity of 139 fb$^{-1}$. 
  Observed data are found to be in agreement with SM background expectations. Results are interpreted in terms of exclusion limits on the masses of the particles
  considered or on the branching ratios associated with the decays. Analyses are employing new techniques, with machine learning algorithms 
  increasingly being used. Sensitity is greatly improved, or set for the first time, with respect to the previous published results in all the considered SUSY scenarios.


\section*{References}
  \begin{thebibliography}{99}
    \bibitem{atlas} ATLAS Collaboration, \href{https://dx.doi.org/10.1088/1748-0221/3/08/S08003}{\Journal{\it JINST}{3}{S08003}{2008}}
    \bibitem{RPVmulti-j} ATLAS Collaboration, \href{https://link.springer.com/article/10.1140/epjc/s10052-021-09761-x}{\Journal{\it \EPJC}{81}{1023}{2021}}
    \bibitem{bRPV} ATLAS Collaboration, \href{https://atlas.web.cern.ch/Atlas/GROUPS/PHYSICS/CONFNOTES/ATLAS-CONF-2022-057}{\textit{ATLAS-CONF-2022-057}}
    \bibitem{LQ} ATLAS Collaboration, \href{https://journals.aps.org/prd/abstract/10.1103/PhysRevD.104.112005}{\Journal{\PRD}{104}{112005}{2021}}
    \bibitem{multi-b} ATLAS Collaboration, \textit{CERN-EP-2022-213}, \href{https://arxiv.org/abs/2211.08028}{arXiv:2211.08028 [hep-ex]}
    \bibitem{photon} ATLAS Collaboration, \textit{CERN-EP-2022-012}, \href{https://arxiv.org/abs/2206.06012}{arXiv:2206.06012 [hep-ex]}
    \bibitem{HPhot} ATLAS Collaboration, \href{https://atlas.web.cern.ch/Atlas/GROUPS/PHYSICS/CONFNOTES/ATLAS-CONF-2023-009}{\textit{ATLAS-CONF-2023-009}}
    \bibitem{EWK1L} ATLAS Collaboration, \href{https://atlas.web.cern.ch/Atlas/GROUPS/PHYSICS/CONFNOTES/ATLAS-CONF-2022-059}{\textit{ATLAS-CONF-2022-059}}
    \bibitem{staus} ATLAS Collaboration, \href{https://atlas.web.cern.ch/Atlas/GROUPS/PHYSICS/CONFNOTES/ATLAS-CONF-2022-042}{\textit{ATLAS-CONF-2022-042}}
    \bibitem{summary} ATLAS Collaboration, \href{https://atlas.web.cern.ch/Atlas/GROUPS/PHYSICS/PUBNOTES/ATL-PHYS-PUB-2023-005}{\textit{ATL-PHYS-PUB-2023-005}}
  \end{thebibliography}
\end{document}